\documentclass[epj]{svjourmod}

\usepackage{graphicx,psfrag}
\usepackage{times}
\usepackage{color}
\usepackage{amsfonts,amssymb,stmaryrd,latexsym,amsmath}

\allowdisplaybreaks[1]

\newcommand{\beq}{\begin{equation}}
\newcommand{\eeq}{\end{equation}}

\DeclareMathOperator{\Si}{Si}
\DeclareMathOperator{\Ci}{Ci}
\DeclareMathOperator{\Vp}{V.p.}
\DeclareMathOperator{\Real}{Re}
\DeclareMathOperator{\Imag}{Im}

\begin{document}

\title{Does the multiple-scattering series in the pion--deuteron scattering actually converge?}
\titlerunning{Does the multiple-scattering series in the pion--deuteron scattering actually converge?}

\author{A.~E.~Kudryavtsev\inst{1} \and A.~I.~Romanov\inst{1,2} \and V.~A.~Gani\inst{1,3}}
\authorrunning{A.~E.~Kudryavtsev {\it et al.}}

\institute{State Scientific Center Institute for Theoretical and Experimental Physics, Moscow, Russia \and
Department of General Physics, National Research Nuclear University MEPhI, Moscow, Russia \and
Department of Mathematics, National Research Nuclear University MEPhI, Moscow, Russia}

\date{
}

\abstract{It is demonstrated that the well-known answer for the multiple-scattering series (MSS) for a light particle interacting to a pair of static nucleons, calculated in the
Fixed Centers Approximation (FCA), works well for a wide region of the two-body complex scattering length $a$. However, this approach is not applicable
in a narrow region surrounding the real positive $a$ half-axis, where the MSS does not converge. Simultaneously, for real positive $a$'s
the 3-body system forms an infinite set of bound states.}


\maketitle

\section{Introduction}

The multiple-scattering series (MSS) has played a prominent role in the study of meson-
nucleus interactions. The sum of all rescattering terms where meson scatters back and 
forth between a pair of static nucleons, i.e., in the Fixed Centers Approximation (FCA) was derived 
by Foldy \cite{Foldy45} in 1945. It was applied to the $\pi d$-scattering by Br\"{u}ckner \cite{Bruckner53} in 1953.
According to Br\"{u}ckner, the $\pi d$-scattering amplitude is the FCA amplitude averaged over the deuteron wave function:
\begin{equation}
\label{1_1}
F_{\pi d}=\int |\psi_d({{\bf r}})|^2 \frac{f_1+f_2+2\frac{f_1f_2}{r}e^{ikr}}{1-\frac{f_1f_2}{r^2}e^{2ikr}}d{\bf r}.
\end{equation}
Here $f_1 (f_2)$ are the $\pi N$ scattering amplitudes on the first (second) nucleon of the deuteron
and $\psi_d({\bf r})$ is the deuteron wave function, $\int |\psi_d({\bf r})|^2d{\bf r}=1$. Expression (\ref{1_1})
was first obtained in \cite{KolKud72} by summing the set of the multiple-scattering diagrams.
More recently the terms of the MSS have been discussed in the context of the effective-field-theory (EFT) treatment
of $\pi$-nucleus scattering. The first EFT calculation of the $\pi d$ scattering amplitude was 
performed by Weinberg \cite{Wein92}. Different aspects of the MSS in EFT formalisms have been refined
in the next twenty years, see, e.g., \cite{refs}. Expression (\ref{1_1}) for $F_{\pi d}$
in an EFT framework was first recreated in \cite{Rus}.
In our recent publication \cite{Baru12} we discussed from the point of view of an EFT
the fact that, if the meson-nucleon scattering is approximated by the scattering length,
the individual terms of the series are divergent and enhanced
with respect to the straightforward expectation from the chiral perturbation theory ($\chi$PT).
We also showed in that work that the divergences cancel if the series is resummed.

In the present paper we make an attempt to analyse the status of the resummed expression for
the $\pi d$ scattering amplitude (\ref{1_1}). Considering the case of equal amplitudes $f_1=f_2=a$, we
get the following expression for the resummed MSS amplitude $F(a)$ in the coordinate space at zero energy:

$$F(a)=F^{(1)}(a)+F^{(2)}(a),$$
where $F^{(1)}(a)=2a$ is the single scattering contribution and

\begin{equation}\label{1_2}
F^{(2)}(a)=2a^2\int |\psi_d({\bf r})|^2 \frac{1}{r-a}d{\bf r}.
\end{equation}
Here $F^{(2)}(a)$ is the sum of the double-scattering term $f^{(2)}(a)$, the triple-scattering term $f^{(3)}(a)$
and so on:

$$F^{(2)}(a)=f^{(2)}(a)+f^{(3)}(a)+...+f^{(n)}(a)+...,$$
where

$$f^{(n)}(a)=2a^n\int |\psi_d({\bf r})|^2\frac{d{\bf r}}{r^{n-1}}\,.$$
Note that we define the scattering length as $a=f(k=0)$,
so the case of $a>0$ corresponds to effective attraction in the $\pi N$-subsystem.

In our previous paper \cite{Baru12} we concentrated mainly on the case of
the negative scattering length, $a<0$. In this case integral (\ref{1_2}) for the amplitude $F^{(2)}(a)$ is well defined.
However, in the case of attraction ($a>0$) the expression for $F^{(2)}(a)$ (\ref{1_2})
becomes formally divergent and one needs additional information
as to how to deal with the pole singularity in Eq. (\ref{1_2}). The first step
towards solving this problem has been done in \cite{Baru12}.
We showed there that the Fourier transform $P(Q)$ of the function $(r-a)^{-1}$ that enters the integral in (\ref{1_2}),
\begin{equation}
P(Q)=\int e^{i{\bf Q}{\bf r}}\frac{1}{r-a}d{\bf r}=\frac{4\pi}{Q^2}+\frac{4\pi a}{Q}f(-aQ),
\end{equation}
where
\begin{equation}
f(y)=\int\limits_0^{+\infty}\frac{\sin x}{x+y}dx=\Ci(y)\sin y+\left(\frac{\pi}{2}-\Si(y)\right)\cos y
\end{equation}
with $\Si(y)$ and $\Ci(y)$ being the sine and cosine integral functions, acquires a cut from $aQ=0$ to $aQ=+\infty$,
due to the cosine integral having a branch point at $y=0$.
It means that at $a>0$ the function $P(Q)$ becomes complex, i.e., it has a nonzero imaginary part.
Returning to the integral representation for $F^{(2)}(a)$ (\ref{1_2}),
we conclude that $F^{(2)}(a)$ has to be considered as an analytic function of the complex parameter
$a=a_1+ia_2$. This leads us to define the amplitude $F^{(2)}(a)$ for real and positive $a>0$
as the limit of $F^{(2)}(a_1+ia_2)$ as $a_2\rightarrow 0$. This prescription
gives for $F^{(2)}(a)$ at $a>0$:
\begin{equation}\label{1_3}
F^{(2)}(a)=8\pi a^2\ \Vp\int |\psi_d(r)|^2\frac{r^2 dr}{r-a}+i\cdot 8\pi^2a^4\psi_d^2(a).
\end{equation}

According to (\ref{1_3}), the $\pi d$ scattering length is complex even
when no inelastic channels are present.
It might mean that the 3-body meson-deuteron scattering problem is not well defined if the meson-nucleon off-shell
amplitude is approximated by the constant scattering length. To check this hypothesis, we can try to find
a representation of the sum of the MSS terms different from that given by Eq. (\ref{1_2}).

Our paper is structured as follows.
In Section 2 we calculate and resum the MSS in the momentum space.
In contrast to what we had in the coordinate space, here we get a reasonable
answer for the amplitude $^m\! F^{(2)}(a)$. However, this answer still reflects the presence of the singularity in 
the expression for $F^{(2)}(a)$ (\ref{1_2}). The resulting solution in the momentum space, obtained numerically,
appears to be an oscillating function of the size of the discretization grid.
To finally decode the answer, we need to introduce a finite range into the elementary $\pi N$-amplitude.
This procedure allows us to sum all the MSS terms and to get finite answers in both the coordinate and
the momentum space representations and to perform a comparison of the results. This is done
in Section 3. In Section 4 we discuss the applicability of both representations of $F^{(2)}(a)$
to the cases of complex and real negative scattering lengths $a$. In the Conclusions we
summarise the results of our research.
In particular, we discuss the physical reasons
behind the final answer for the MSS amplitude becoming ill-defined when the range of the elementary $\pi N$ amplitude goes to zero.

\section{The basic equation for the MSS in the momentum space}

Using the Feynman diagram technique one can easily obtain (see, e.g., \cite{KolKud72,BaruKud97}) 
3-dimensional expressions for the multiple-scattering amplitudes $f^{(n)}(a)$ in the FCA:

\begin{equation}\label{2_1}
f^{(n)}(a)=\int\frac{\varphi_d({\bf p})}{(2\pi)^3}\Sigma^{(n)}({\bf p},{\bf p}\,')
\frac{\varphi_d({\bf p}\,')}{(2\pi)^3}d{\bf p}d{\bf p}\,',\ n=2,3,...,
\end{equation}
where $\varphi_d({\bf p})$ is the deuteron wave function in the momentum space, normalized such that
$\int |\varphi_d({\bf p})|^2d{\bf p}=(2\pi)^3$, and
\begin{equation}
\begin{split}
  \Sigma^{(2)}({\bf p},{\bf p}\,')=&2a^2\frac{4\pi}{({\bf p}-{\bf p}\,')^2+{\varkappa}^2},{}\\
  \Sigma^{(3)}({\bf p},{\bf p}\,')=&2a^3\int\frac{d{\bf s}}{(2\pi)^3}
  \frac{4\pi}{({\bf p}-{\bf s})^2+{\varkappa}^2}\frac{4\pi}{({\bf s}-{\bf p}\,')^2+{\varkappa}^2},{}\\
  \Sigma^{(4)}({\bf p},{\bf p}\,')=&
\nonumber
\end{split}
\end{equation}
\begin{equation}\label{2_2}
\begin{split}
  2a^4\int\frac{d{\bf s}d{\bf t}}{(2\pi)^6}
  \frac{4\pi}{({\bf p}-{\bf s})^2+{\varkappa}^2}\frac{4\pi}{({\bf s}-{\bf t})^2+{\varkappa}^2}
  \frac{4\pi}{({\bf t}-{\bf p}\,')^2+{\varkappa}^2}\,,
\end{split}
\end{equation}
and so on.
To avoid the IR singularities in the integrals for $\Sigma^{(n)}({\bf p},{\bf p}\,')$, we introduce
the IR cutoff $\varkappa$ everywhere in the denominators of the pionic propagators.
Taking into account only the leading S-wave part of the deuteron wave function,
one can perform all the angular integrations in (\ref{2_1}), resulting in
\begin{equation}
\begin{split}
  f^{(2)}(a)=2a^2\frac{1}{4\pi^3}&\int\limits_0^{+\infty}p\varphi_d(p)\{\ln_{\varkappa}(p,p')\}\varphi_d(p')p'dpdp',{}\\
  f^{(3)}(a)=2a^3\frac{1}{4\pi^3}\times&
\nonumber
\end{split}
\end{equation}
\begin{equation}\label{2_3}
\begin{split}
\int\limits_0^{+\infty}p\varphi_d(p)\left(\int\limits_0^{+\infty}\frac{ds}{2\pi}
  \ln_{\varkappa}(p,s)\ln_{\varkappa}(s,p')\right)\varphi_d(p')p'dpdp',
  \end{split}
\end{equation}
and so on, where we denote
$$\ln_{\varkappa}(p,p')=\ln\frac{(p+p')^2+\varkappa^2}{(p-p')^2+\varkappa^2}.$$
The answer for the full MSS amplitude $^m\! F^{(2)}(a)=\sum\limits_{n=2}^{+\infty}f^{(n)}(a)$
in the momentum space representation then reads:
\begin{equation}\label{2_4}
  ^m\! F^{(2)}(a)=2a^2\frac{1}{4\pi^3}\int\limits_0^{+\infty}p\varphi_d(p)R(p,p')\varphi_d(p')p'dpdp',
\end{equation}
where the function $R(p,p')$ is a solution of the integral equation:
\begin{equation}\label{2_5}
  R(p,p')=\ln_{\varkappa}(p,p')+a\int\limits_0^{+\infty}\frac{ds}{2\pi}\ln_{\varkappa}(p,s)R(s,p').
\end{equation}
It is convenient to introduce dimensionless variables $x,y,z$ and a parameter $L$:

$$p=p_0x, \enskip p'=p_0y, \enskip s=p_0z; \enskip \varkappa=p_0L.$$

To solve equation (\ref{2_5}) numerically we introduce the upper limit of the integration (regularization),
$\Lambda=Np_0$. We take the parameter $N$ to be an integer.
To satisfy the boundary conditions $R(p,0)\equiv R(0,p')\equiv 0$,
the parameters $p_0, \varkappa$ and $\Lambda$ have to satisfy $p_0\ll\varkappa\ll\Lambda$
(or $1\ll L \ll N$). Now we introduce a dimensionless integer-valued grid of size $N$ with the step 1.
The discrete version of our basic equation (\ref{2_5}) reads:
\begin{equation}\label{2_6}
  R_N(i,j)=\ln_L(i,j)+\epsilon\sum\limits_{k=1}^N\ln_L(i,k)R_N(k,j),
\end{equation}
where we denote $R_N(i,j)\equiv R_N(x_i,y_j)$,
$\ln_L(i,j)\equiv \ln_L(x_i,y_j)$, $x_i=i-\frac{1}{2}, y_j=j-\frac{1}{2}, 1\leq i,j\leq N$ and
$\epsilon=\displaystyle\frac{ap_0}{2\pi}$. The solution $R_N(i,j)$ of equation (\ref{2_6}) depends on the size $N$ of the grid, and is a matrix
$N\times N$.

\begin{figure}
\centering
\includegraphics[width=\linewidth]{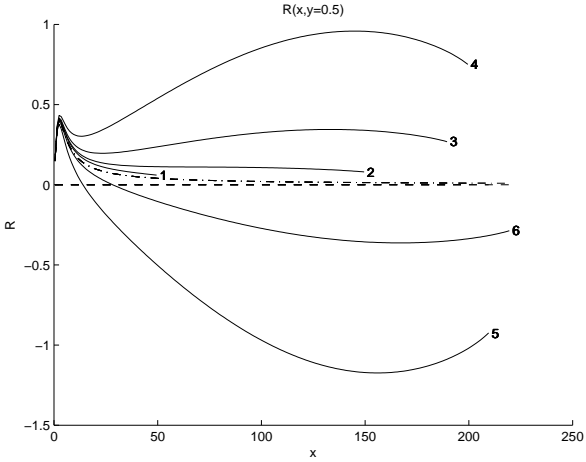}
\caption{The dependence of the function $R(x,y)$ on $x$ at $y=0.5$ for some typical cutoffs $N$ (here $N$ is the size of the grid).
The solid curve 1 corresponds to $N=50$; 2 -- to $N=150$; 3 -- to $N=190$; 4 -- to $N=200$; 5 -- to $N=210$; 6 -- to $N=220$.
The first critical point $N_{crit}^{(1)}=205$. The dashed-dotted curve shows the Born perturbative term
$R_0(x,y)=\ln\frac{(x+y)^2+L^2}{(x-y)^2+L^2}$, also at $y=0.5$. The parameters used in the calculation:
$a=10^{-3}\ \mathrm{MeV}^{-1}$, $p_0=10\ \mathrm{MeV}$, $L=2.5$.}
\label{Fig01}
\end{figure}

\begin{figure}
\centering
\includegraphics[width=\linewidth]{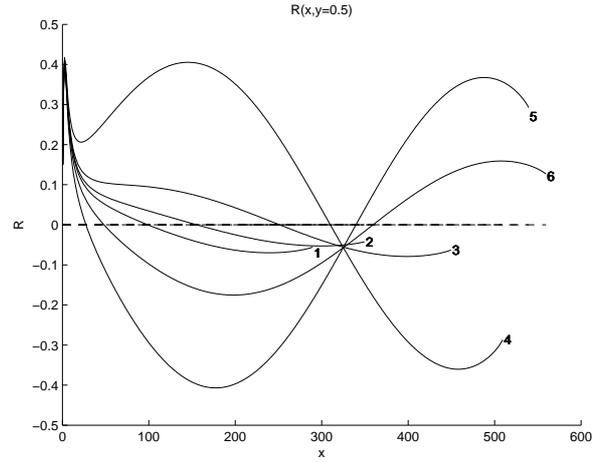}
\caption{The same as in Fig.~\ref{Fig01} but for larger $N$. Curve 1 corresponds to $N=290$; 2 -- to $N=350$; 3 -- to $N=450$; 4 -- to $N=510$; 5 -- to $N=540$; 6 -- to $N=560$.
The second critical point $N_{crit}^{(2)}=525$.}
\label{Fig02}
\end{figure}

Solving Eq. (\ref{2_6}) numerically, we found that $R_N(i,j)$ nontrivially depends on $N$.\footnote{Note that special
efforts were performed to check stability of the results of our numerical procedure. In particular, we varied scale
$p_0\to\widetilde{p}_0=p_0/n$ and $N\to\widetilde{N}=N\cdot n$ ($n=2,3,4$) and found that numerical results survive.}
Some examples of the solution $R_N(x,y)$ at different $N$ are presented in Figs.~\ref{Fig01}, \ref{Fig02}.
The evolution
of $R_N(x,y=0.5)$ as a function of $N$ is the following. At small $N$ the exact solution behaves similarly
to the function $R_0(x,y)$, which is the leading perturbative approximation for $R_N(x,y)$ in terms of the parameter $\epsilon$,
$$R_0(x,y)=\ln\frac{(x+y)^2+L^2}{(x-y)^2+L^2}.$$
The function $R_0(x,y=0.5)$ is pictured in Fig.~\ref{Fig01} by the dashed-dotted line. We see that
the solid curves 1 and 2 in Fig.~\ref{Fig01} behave similarly to $R_0(x,y)$. However, with growing
$N$ the solution $R_N(x,0.5)$ deviates sharply from $R_0(x,0.5)$.
At $N$ near $N_{crit}^{(1)}=205$, the solution blows up to the positive infinity,
then changes its sign to become the negative infinity
and continues its evolution as $N$ increases (compare curves 4, 5 and 6 of Fig.~\ref{Fig01}).
Simultaneously, at $N>N_{crit}^{(1)}$ solution $R_N(x,0.5)$ develops a node at small $x$.
This node moves to the right with the growth of $N$ (compare curves 5 and 6 of Fig.~\ref{Fig01}).

The evolution of $R_N(x,0.5)$ with $N$ further increasing is shown in Fig.~\ref{Fig02}.
The solution blows up again at $N$ near the second critical point $N_{crit}^{(2)}=525$ and the story repeats.
At $N>N_{crit}^{(2)}$, the second node of $R_N(x,0.5)$ emerges, see curves 5 and 6 in Fig.~\ref{Fig02}.

\begin{figure}
\centering
\includegraphics[width=\linewidth]{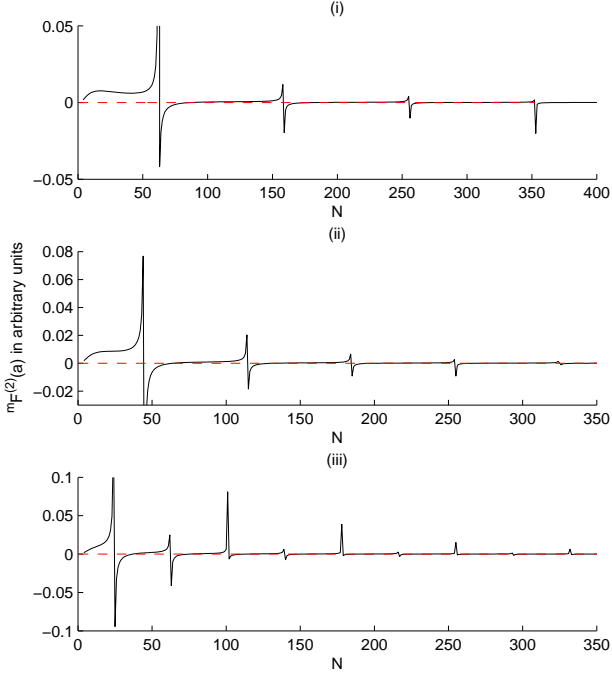}
\caption{The dependence of the scattering amplitude $^m\! F^{(2)}(a)$ on the grid size $N$.
The calculation is performed with the following parameters: i) $a=0.0035\ \mathrm{MeV}^{-1}$; ii) $a=0.005\ \mathrm{MeV}^{-1}$; iii) $a=0.01\ \mathrm{MeV}^{-1}$.}
\label{Fig03}
\end{figure}

\begin{figure}
\centering
\includegraphics[width=\linewidth]{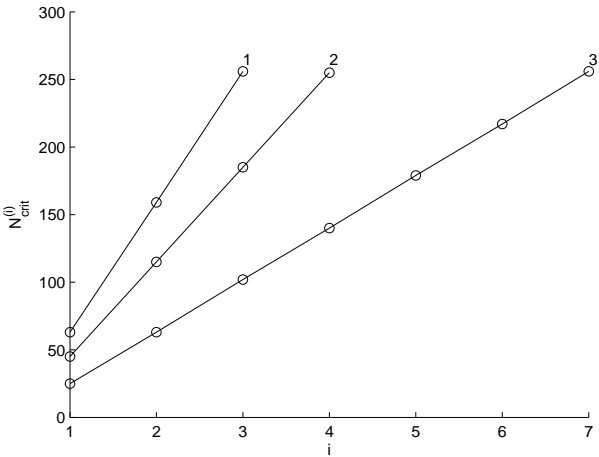}
\caption{The dependence of pole positions $N_{crit}^{(i)}$ on $i$ for different values of $a$.
Line 1 corresponds to $a=0.0035\ \mathrm{MeV}^{-1}$, 2 -- to $a=0.005\ \mathrm{MeV}^{-1}$ and 3 -- to $a=0.01\ \mathrm{MeV}^{-1}$.}
\label{Fig04}
\end{figure}

The further observed evolution of $R_N(x,0.5)$ with $N$ appears to be periodic or cyclic --- there is an infinite
series of critical points ${N_{crit}^{(i)}}$, and the solution $R(x,0.5)$ blows up in the vicinity of each $N_{crit}^{(i)}$.
This behavior of the amplitude $^m\! F_N^{(2)}$ versus $N$ is demonstrated in Fig.~\ref{Fig03} for some selected scattering lengths $a$.
Within the accuracy of calculation the distance between two adjacent critical points $(N_{crit}^{(i+1)}-N_{crit}^{(i)})$
was found to be independent of $i$. This observation means the dependence of $N_{crit}^{(i)}$ on $i$ is linear,
which is demonstrated in Fig.~\ref{Fig04} for different scattering lengths $a$.

Note that in the vicinity of any $N_{crit}^{(i)}$ the amplitude $^m\! F_N^{(2)}(a)$ behaves similarly
to the situation when a new bound state is created at threshold: it grows to infinity and then changes its sign.
We can argue that what we observe in our numerical calculation is an evidence
that, in the limit $\Lambda\rightarrow\infty$, the 3-body system we are studying has an infinite number of bound states.
To prove this hypothesis conclusively, we need to search for the source
of creation of the series of bound states in the $\pi d$-system. For this purpose we are going to introduce a finite range $r_0$
of the force (which is equivalent to a formfactor) for the elementary $\pi N$-system
and to study the dependence of the amplitude $F^{(2)}(a)$ on $r_0$. This scenario will be the subject of the next section.

\section{Introducing a formfactor into the elementary $\pi N$-amplitude}

The elementary $\pi N$-interaction potential is, in general,
characterized by a certain finite range of force and hence the $t$-matrix is momentum dependent.
The impact of the inclusion of a formfactor in the function $F^{(2)}(a)$
in the coordinate space was already discussed in \cite{Baru12}. Here we want to introduce
formfactor in the momentum space and to compare the results of both approaches.

Consider the full off-shell $\pi N$ amplitude in the separable form:
$$f(p,p';E=0)=ag(p)g(p'),$$
where we take
$$g(p)=\frac{M}{\sqrt{M^2+p^2}}\,.$$
The limit of the zero-range $\pi N$-interaction corresponds to $M$ going to infinity.

In the coordinate space, the introduction of this formfactor leads to the following modification
of the expression for $F^{(2)}(a)$ (\ref{1_2}):
\begin{equation}\label{3_1}
\begin{split}
  F_M^{(2)}&(a)=2a^2\frac{M^2}{M^2-\varkappa^2}\times\\
&\int |\psi_d({\bf r})|^2 \frac{e^{-\varkappa r}-e^{-Mr}}{r-\left( \frac{M^2}{M^2-\varkappa^2} \right)a(e^{-\varkappa r}-e^{-Mr})}d{\bf r}.
\end{split}
\end{equation}
In the limit $\varkappa \rightarrow 0$ this expression coincides with that discussed in \cite{Baru12}.
In contrast to what we had in the point-like limit (\ref{1_2}), this expression for $F_M^{(2)}(a)$ is well defined in the
region $\displaystyle Ma<1+\frac{\varkappa}{M}$.
At $\displaystyle Ma>1+\frac{\varkappa}{M}$ the pole singularity in the denominator emerges again.

In the momentum space, the expression for the amplitude $^m\! F_M^{(2)}(a)$ with the formfactor $g(p)$ reads:
\begin{equation}\label{3_2}
\begin{split}
  ^m\! F_M^{(2)}(a)=&2a^2\frac{1}{4\pi^3}\left(\frac{M^2}{M^2-\varkappa^2}\right)\times\\
&\int\limits_0^{+\infty} p\varphi_d(p)R_M(p,p')\varphi_d(p')p'dpdp',
\end{split}
\end{equation}
where $R_M(p,p')$ is the solution of the following integral equation (cf.\ Eq.~(\ref{2_5})):
\begin{equation}\label{3_3}
\begin{split}
  R_M(p,p')=&\ln_{\varkappa,M}(p,p')+\\
&a\frac{M^2}{M^2-\varkappa^2}\int\limits_0^{+\infty}\frac{ds}{2\pi}\ln_{\varkappa,M}(p,s)R_M(s,p'),
\end{split}
\end{equation}
where
$$\ln_{\varkappa,M}(p,p')=\ln_{\varkappa}(p,p')-\ln_M(p,p').$$

Comparing the functions $F_M^{(2)}(a)$ and $^m\! F_M^{(2)}(a)$,
calculated in both the coordinate and
momentum space representations, we find that they practically coincide for the values of
the parameter $M\leq M_0$, where $M_0$ is determined from the equation:
\begin{equation}\label{3_4}
  M_0 a=1+\frac{\varkappa}{M_0}.
\end{equation}
For $M>M_0$ the function $F_M^{(2)}(a)$ in the coordinate representation is formally divergent similar to the point-like case discussed in the previous section.

\begin{figure}[h]
\centering
\includegraphics[width=\linewidth]{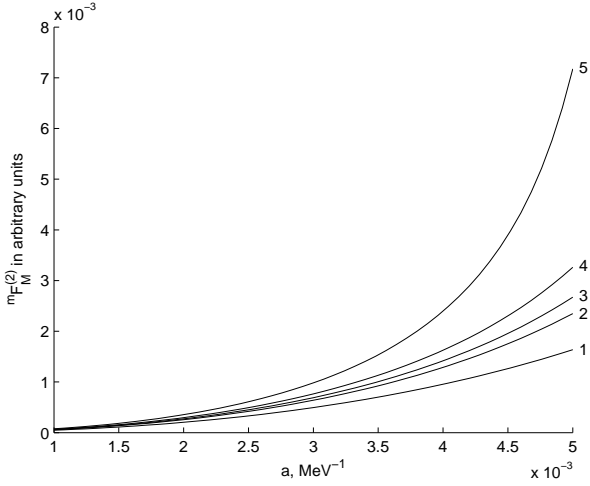}
\caption{The function $^m\! F_M^{(2)}(a)$ versus $a$ for different values of $M$.
Curve 1 corresponds to $M=150\ \mathrm{MeV}$; 2 -- to $M=220\ \mathrm{MeV}$; 3 -- to $M=250\ \mathrm{MeV}$; 4 -- to $M=300\ \mathrm{MeV}$; 5 -- to $M=500\ \mathrm{MeV}$.}
\label{Fig05}
\end{figure}

\begin{figure}
\centering
\includegraphics[width=\linewidth]{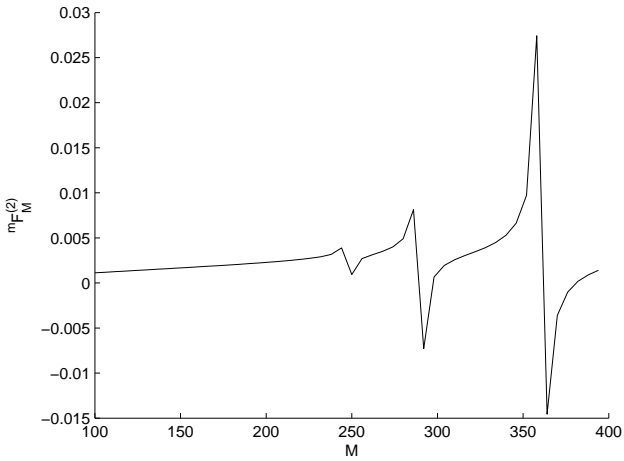}
\caption{The function $^m\! F_M^{(2)}(a)$ versus $M$ for $a=0.005\ \mathrm{MeV}^{-1}$, $M_0=222\ \mathrm{MeV}$.}
\label{Fig06}
\end{figure}

\begin{figure}
\centering
\includegraphics[width=\linewidth]{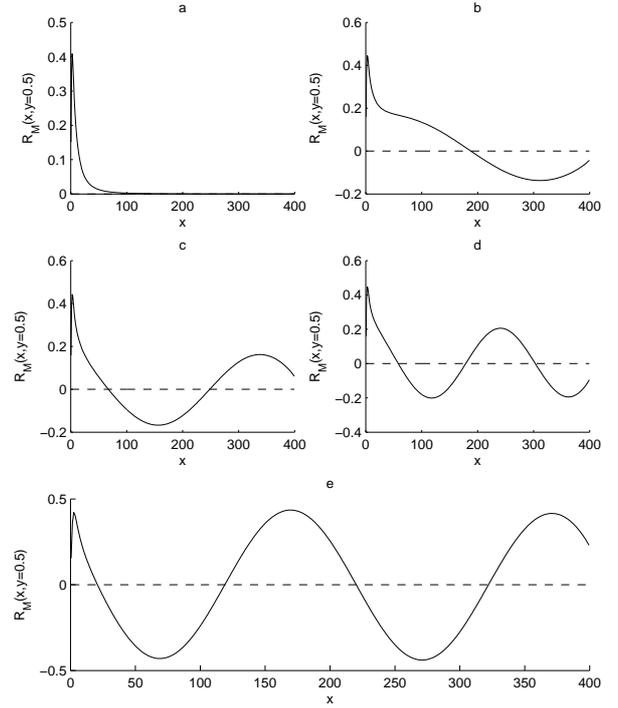}
\caption{The function $R_M(x,y=0.5)$ versus $x$ at different typical $M$: a) $M=150\ \mathrm{MeV}$; b) $M=260\ \mathrm{MeV}$; c) $M=280\ \mathrm{MeV}$; d) $M=330\ \mathrm{MeV}$; e) $M=380\ \mathrm{MeV}$.}
\label{Fig07}
\end{figure}

The function $^m\! F_M^{(2)}(a)$, calculated in the momentum space according to Eq.~(\ref{3_2}), is well defined also for $M>M_0$.
Some examples of the solution for $^m\! F_M^{(2)}(a)$ in the momentum space versus $a$ are demonstrated in Fig.~\ref{Fig05}
for some different values of $M$. All calculations were carried out with the cut-off parameter $\Lambda=2500$ MeV.
For $M>M_0$ these functions depend crucially on $M$.
The amplitude $^m\! F_M^{(2)}(a)$, calculated in the momentum space representation at
$a=5\cdot 10^{-3} \mathrm{MeV}^{-1}$ as a function of $M$, is shown in Fig.~\ref{Fig06}.
One can see the set of critical points $M_{crit}^{(i)}$ which appear as spikes.
All the singular points $M=M_{crit}^{(i)}$ are located to the right from the point $M=M_0$ (\ref{3_4}),
$M_{crit}^{(i)}>M_0$. At $M<M_0$ the amplitude $^m\! F_M^{(2)}(a)$ is smooth, positive and finite.
For $M\rightarrow M_{crit}^{(1)}$ \quad $^m\! F_M^{(2)}(a)$ goes to $+\infty$, then changes its sign and continues its further evolution
with $M$. A similar behavior is seen near the second critical point $M_{crit}^{(2)}$.
The function $R_M(x,y=0.5)$ is nodeless for $M$ being in the interval
$M\in(\varkappa,M_{crit}^{(1)})$, while it develops its first node in the interval $M\in(M_{crit}^{(1)},M_{crit}^{(2)})$.
Some typical pictures for the function $R_M(x,y=0.5)$ are demonstrated in Fig.~\ref{Fig07}. The further evolution of 
$R_M(x,y=0.5)$ with $M$ also appears to be cyclic.

We conclude that for $M<M_0$ the 3-body system under our study
appears to be in an attractive but perturbative regime without forming any bound state.
At $M=M_{crit}^{(1)}$ the attraction is getting strong just enough to form the first bound state.
The amplitude $F_M^{(2)}(a)$ for $M>M_{crit}^{(1)}$ becomes negative, which
corresponds to the effective repulsion. With the further
growth of $M$ this picture repeats periodically. In the limit $M\rightarrow +\infty$
an infinite set of bound states in the $\pi d$-system is formed. Hence we conclude that the perturbative approach
is not applicable in this extremely nonperturbative case of positive scattering length $a$.
Looking again at Fig.~\ref{Fig06}, we also conclude that the function $^m\! F_M^{(2)}(a)$
behaves akin to $\tan M$, i.e., periodically oscillates from minus to plus infinity.
It means that this function has no definite limit as $M$ goes to infinity, i.e., in the point-like limit. This observation is in line
with the fact that the function $F_M^{(2)}(a)$ in the coordinate representation
is not well-defined for $M>M_0$. On the other hand, treating $F^{(2)}(a)$ as an analytic function of
$a=a_1+ia_2$, we come to expression (\ref{1_3}).

Thus, the straightforward solution of the integral equation for the function $R_M(p,p')$
at real positive values of scattering length $a$ tells us
that the function $^m\! F^{(2)}(a)=\lim\limits_{M\rightarrow +\infty}^{\phantom{I}} {\phantom{I}} ^m\! F_M^{(2)}(a)$ does not exist.
The analyticity hypothesis for the function $F^{(2)}(a)$, on the other hand,
gives us a unique solution for $F^{(2)}(a)$ (see Eq.~(\ref{1_3})).
The situation looks rather ambiguous. To make the correct choice and to solve the problem
we should study the properties of the function $F^{(2)}(a)$ in the
complex $a$-plane. It will be the subject of discussion in the next section.

\section{Inclusion of absorption}

In the real $\pi d$-system the effect of absorption is not the leading one because both 
elastic $\pi N$-amplitudes ($\pi^-p\rightarrow\pi^-p$ and $\pi^-n\rightarrow\pi^-n$) are real.
The role of absorption is more crucial in the $K^-d$-scattering problem because of
the presence of open inelastic two-body channels $K^-N\rightarrow\pi\Sigma(\pi\Lambda)$.
The FCA is widely used for the $K^-d$ scattering problem, see, e.g., \cite{Kamalov}.
For more recent publications see \cite{Oset} and references therein.

\begin{figure}
\centering
\includegraphics[width=\linewidth]{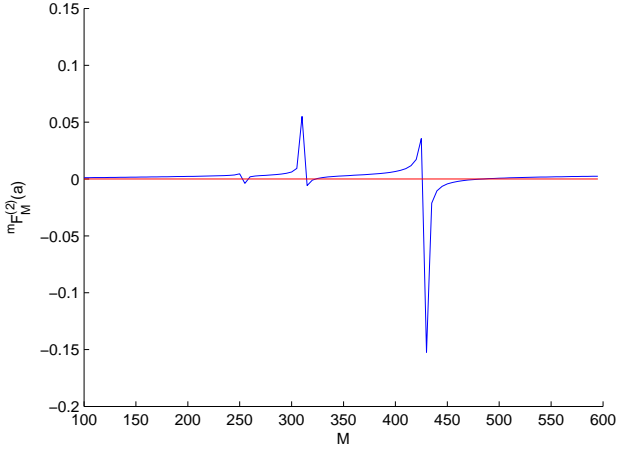}
\caption{The function $^m\! F_M^{(2)}(a)$ for scattering length $a=0.005\ \mathrm{MeV}^{-1}$.}
\label{Fig08}
\end{figure}

\begin{figure}
\centering
\includegraphics[width=\linewidth]{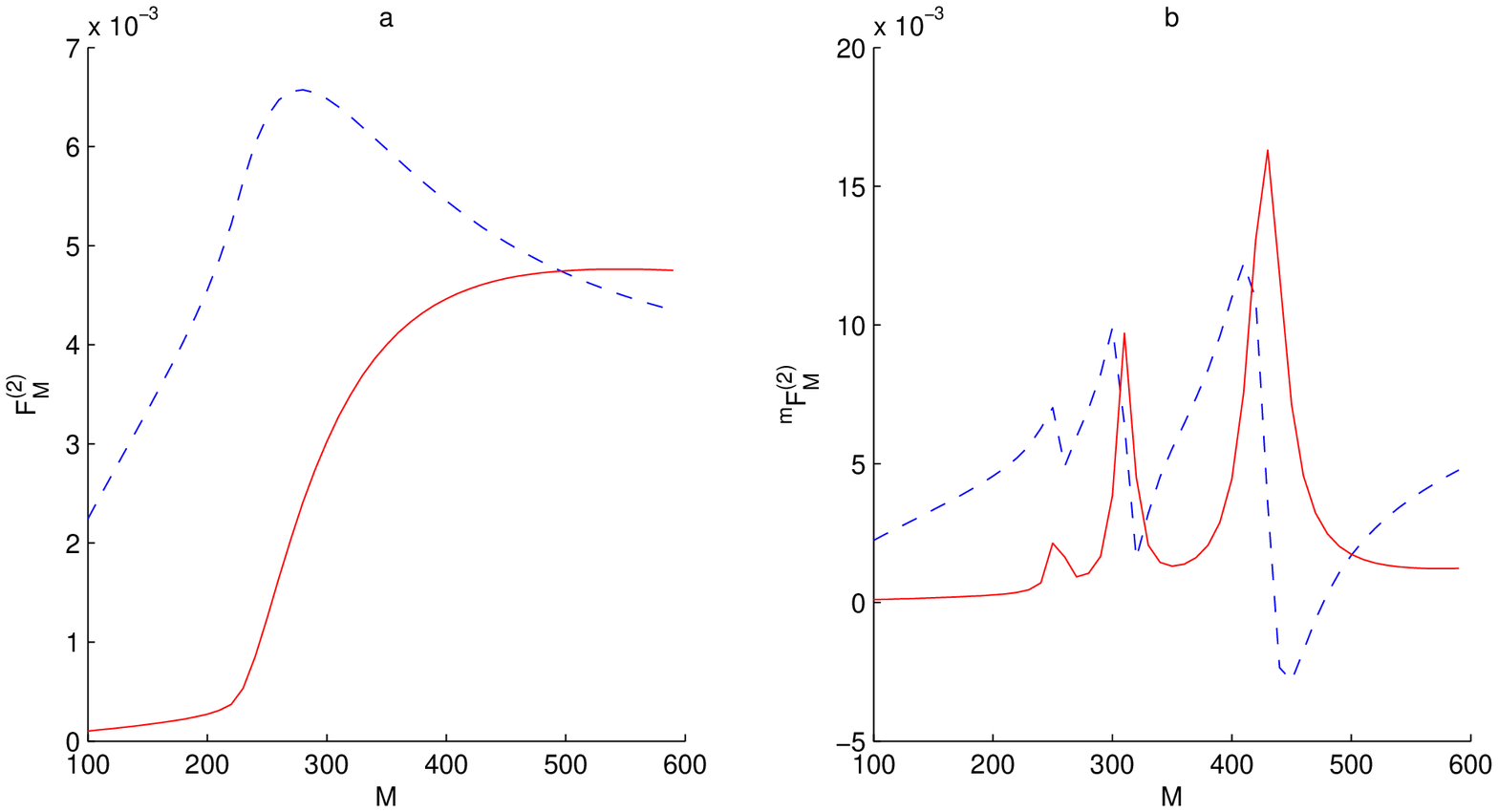}
\caption{Real (dashed curve) and imaginary (solid curve) parts of the functions $F_M^{(2)}(a)$ (plot ({\it a})) and $^m\! F_M^{(2)}(a)$ (plot ({\it b})) for the scattering length $a=0.005+0.0001i\ \mathrm{MeV}^{-1}$.}
\label{Fig09}
\end{figure}

\begin{figure}
\centering
\includegraphics[width=\linewidth]{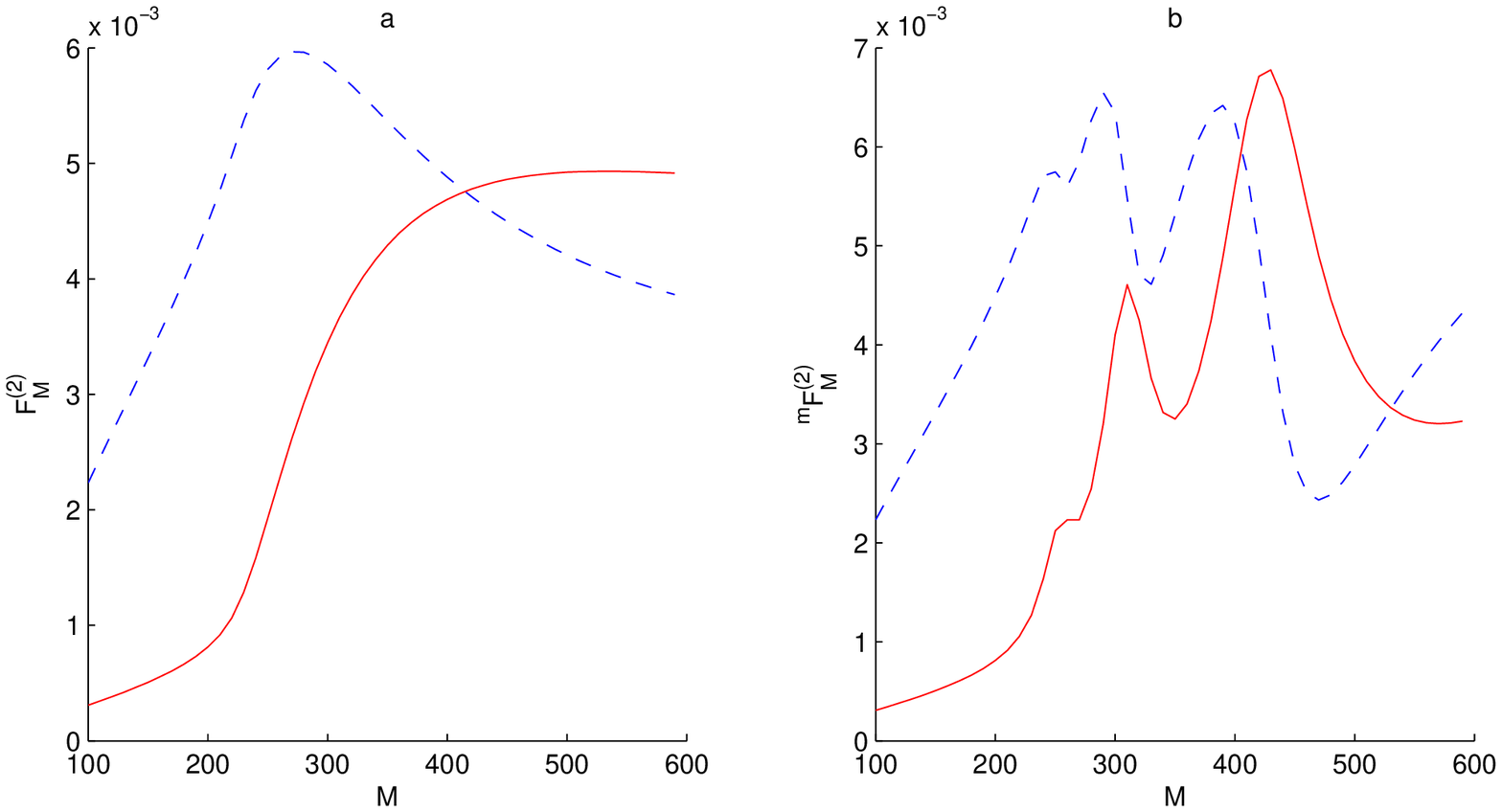}
\caption{Real (dashed curve) and imaginary (solid curve) parts of the functions $F_M^{(2)}(a)$ (plot ({\it a})) and $^m\! F_M^{(2)}(a)$ (plot ({\it b})) for the scattering length $a=0.005+0.0003i\ \mathrm{MeV}^{-1}$.}
\label{Fig10}
\end{figure}

\begin{figure}
\centering
\includegraphics[width=\linewidth]{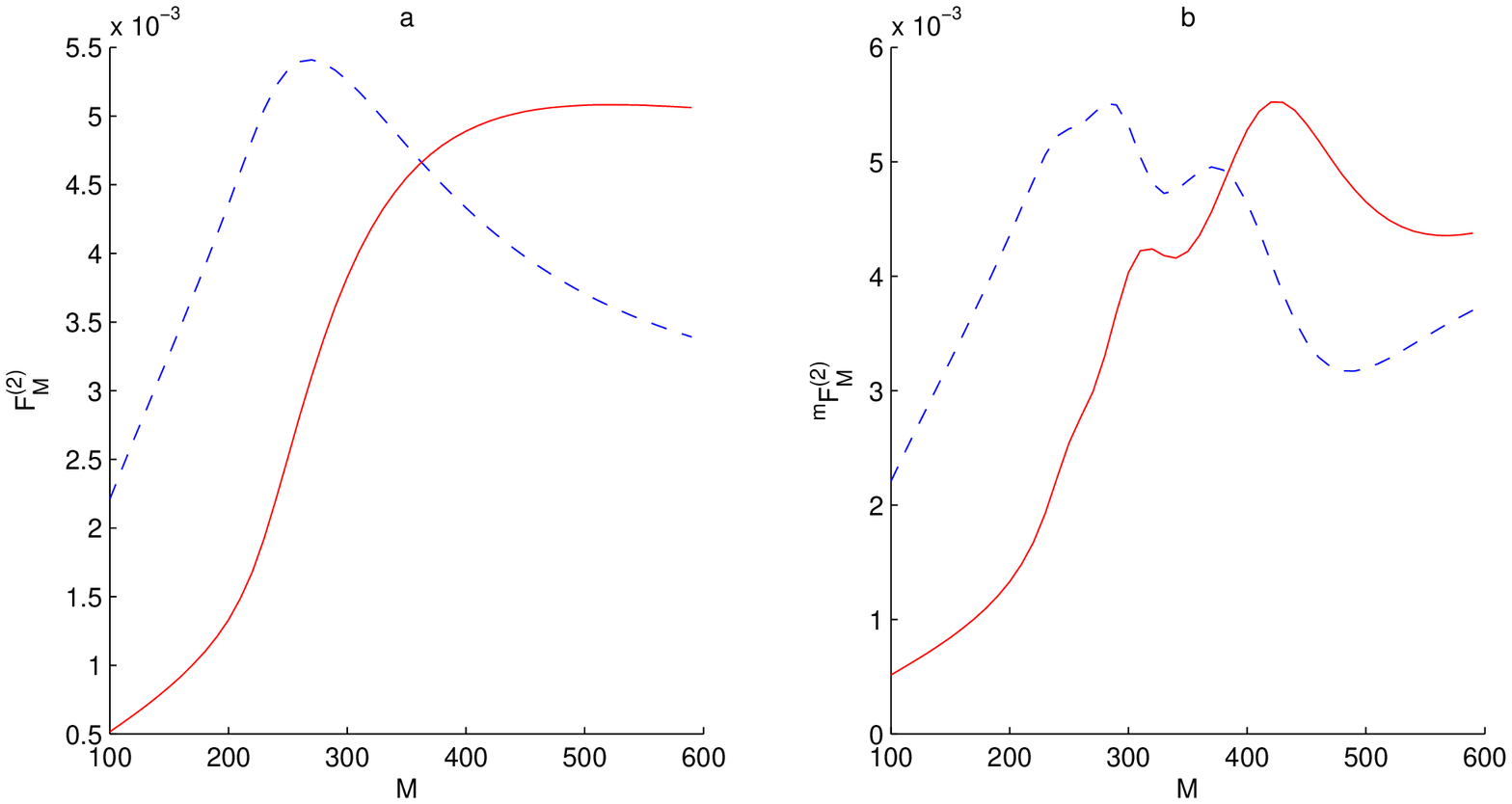}
\caption{Real (dashed curve) and imaginary (solid curve) parts of the functions $F_M^{(2)}(a)$ (plot ({\it a})) and $^m\! F_M^{(2)}(a)$ (plot ({\it b})) for the scattering length $a=0.005+0.0005i\ \mathrm{MeV}^{-1}$.}
\label{Fig11}
\end{figure}

\begin{figure}
\centering
\includegraphics[width=\linewidth]{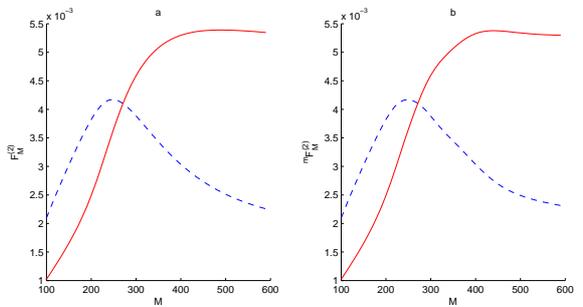}
\caption{Real (dashed curve) and imaginary (solid curve) parts of the functions $F_M^{(2)}(a)$ (plot ({\it a})) and $^m\! F_M^{(2)}(a)$ (plot ({\it b})) for the scattering length $a=0.005+0.001i\ \mathrm{MeV}^{-1}$.}
\label{Fig12}
\end{figure}

Note that expression (\ref{1_2}) for the amplitude $F^{(2)}(a)$ is well defined
only if the imaginary part of the scattering length $a=a_1+ia_2$ differs from zero, $a_2\neq 0$.
On the other hand, we may solve Eq.~(\ref{3_3}) for the function $R_M(p,p')$ considering the parameter $a$
as a complex one. Calculating this function $^m\! F_M^{(2)}(a)$ in the momentum space, one can compare the
results of calculations for $F_M^{(2)}(a)$ in both the momentum and the space representations.
This comparison is demonstrated in Figs.~\ref{Fig08}--\ref{Fig12} for the cut-off parameter in Eq.~(\ref{3_3}) $\Lambda=2500$ MeV.
The real part of the scattering length was fixed, $$a_1=5\cdot 10^{-3} \mathrm{MeV}^{-1}\,,$$
and we varied the imaginary part starting from very small positive $a_2$.
The starting point is Fig.~\ref{Fig08} that reproduces the function $^m\! F_M^{(2)}(a)$ versus $M$
for real and positive $a=a_1=5\cdot 10^{-3} \mathrm{MeV}^{-1}$. The function $^m\! F_M^{(2)}(a)$ in this figure is analogous
to what is shown in Fig.~\ref{Fig06}. We see quite a regular behavior of $^m\! F_M^{(2)}(a)$ at small $M<M_0$
and several singular points (or spikes) at $M>M_0$. The imaginary part of $^m\! F_M^{(2)}(a)$ is equal to zero.

In Fig. \ref{Fig09}{\it b} we draw the real and imaginary parts of the function $^m\! F_M^{(2)}(a)$, calculated in the momentum space
for complex scattering length $a=0.005+0.0001i\, (\mathrm{MeV}^{-1})$. Both $\Real\: ^m\! F_M^{(2)}(a)$ and $\Imag\: ^m\! F_M^{(2)}(a)$ are shown in this figure.
The behavior of the real and imaginary parts of $^m\! F_M^{(2)}(a)$ at $M$ near each of the $M_{crit}^{(i)}$
looks similar to their behavior in the vicinity of a Breit-Wigner resonance.
The next step of the evolution of the function $^m\! F_M^{(2)}(a)$ with increasing $\Imag a=a_2$ is shown in Fig.~\ref{Fig10}{\it b}
($a_2=0.0003\enskip \mathrm{MeV}^{-1}$). Although for this value of $a_2$ the peaks at this figure are still separated,
their corresponding regions already start to overlap. This tendency is better seen in Fig.~\ref{Fig11}{\it b} ($a_2=0.0005\enskip \mathrm{MeV}^{-1}$)
where the isolated peaks are seriously squeezed or deformed. Further evolution of $^m\! F_M^{(2)}(a)$ with
the growth of $a_2$ is demonstrated in Fig.~\ref{Fig12}{\it b}. We see in this figure that isolated
peaks disappear and both the real and imaginary parts of the function $^m\! F_M^{(2)}(a)$ become smooth
functions --- no peaks any more. We observe therefore that if the absorption ($\Imag a$) is not extremely small, i.e., $a_2$
is comparable with $a_1$, all the calculated curves behave quite regularly.

Now it is time to compare these results with what we get in the coordinate
representation, i.e., with the integral representation given by Eq. (\ref{3_1}). In Fig. \ref{Fig09}{\it a} the real
and imaginary parts of $F_M^{(2)}(a)$, calculated according to Eq. (\ref{3_1})
for $a=0.005+0.0001i\, (\mathrm{MeV}^{-1})$, are shown. Both functions look quite regular
and smooth. Note that $\Imag F_M^{(2)}(a)$ is extremely small at small $M<M_0\approx 220$.
This is a consequence of the fact that $\Imag a=a_2$ is taken extremely small. The growth of
$\Imag F_M^{(2)}(a)$ for $M>M_0$ looks like a threshold phenomenon. The origin of this non-negligible
imaginary part is similar to what is given in the r.h.s.\ of Eq.~(\ref{1_3}) for the point-like case.
This contribution may be called the ``unphysical'' part of $\Imag\: ^m\! F_M^{(2)}(a)$.
In the limit $a_2 \rightarrow 0$ and $M\rightarrow\infty$ the function $\Imag F_M^{(2)}(a)$
coincides with the imaginary part given in Eq.~(\ref{1_3}). 
Further evolution of $F_M^{(2)}(a)$ at different values of $a_2$ is demonstrated in Figs.~\ref{Fig10}{\it a} -- \ref{Fig12}{\it a}.
As it is seen, the contribution of the ``physical'' absorption is getting more prominent with
growing $a_2$, which is intuitively clear as absorption on each individual scatterer is getting stronger.

It is instructive to compare Figs.~\ref{Fig09}{\it a} and \ref{Fig09}{\it b},
drawn for the same values of $a_2$. At extremely small $a_2$ they look absolutely different.
The function $^m\! F_M^{(2)}(a)$, calculated in the momentum space for small $a_2$, is an
oscillating function of $M$ with singular points. Because of the oscillating character no definite answer for this function
in the point-like limit $M\rightarrow +\infty$ exists. However, with the growing $a_2$ $^m\! F_M^{(2)}(a)$
is getting smooth and for large $a_2$ the profiles of the functions $F_M^{(2)}(a)$ and $^m\! F_M^{(2)}(a)$
calculated by different methods, coincide with each other (compare, e.g., the curves in Figs.~\ref{Fig12}{\it a} and \ref{Fig12}{\it b}).
Note that at relatively large $a_2$
$$\lim\limits_{M\rightarrow\infty} F_M^{(2)}(a)=F^{(2)}(a),$$
where $F^{(2)}(a)$ is the point-like limit for scattering amplitude given by Eq. (\ref{1_2}).

\begin{figure}
\centering
\includegraphics[width=\linewidth]{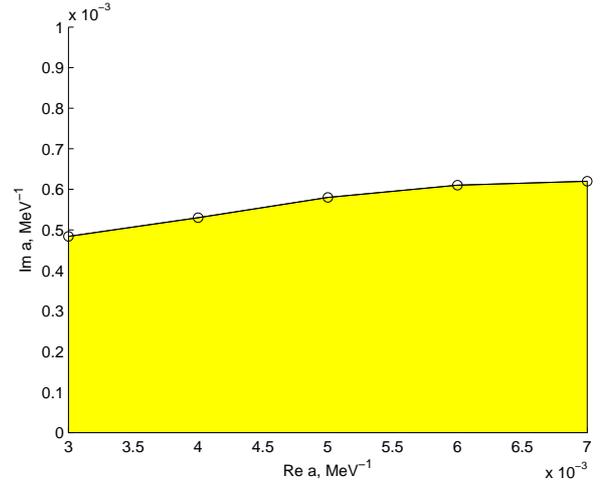}
\caption{The region of complex parameter $a$ in which the functions $F_M^{(2)}(a)$ and $^m\! F_M^{(2)}(a)$, calculated with $\Lambda=2500$ MeV,
don't coincide with accuracy better than 1\%.}
\label{Fig13}
\end{figure}

Concluding this section, we would like to stress that we used the momentum-space
representation to check the validity of the simple coordinate-space representation, given by Eq.~(\ref{1_2}).
We came to the conclusion that almost everywhere in the complex plane of parameter $a$ one can
use Eq.~(\ref{1_2}) to calculate the MSS. However, this equation gives incorrect answer
in a narrow region of complex $a$. This exceptional region is shown in Fig.~\ref{Fig13}
for the particular value $\Lambda=2500$ MeV.
The border of the dashed area is chosen under condition that functions $F_M^{(2)}(a)$ and $^m\! F_M^{(2)}(a)$,
calculated in the two different representations,
coincide with accuracy better than 1\%. Everywhere outside the marked area the FCA given by the simple and useful expression
(\ref{1_2}) is applicable. In the case of real negative $a\leq 0$,
the answers for the amplitudes $F_M^{(2)}(a)$ and $^m\! F_M^{(2)}(a)$ practically coincide identically as well.

Note that the results mentioned above were obtained at $\Lambda=2500$ MeV,
but we have also performed calculations with larger values of $\Lambda$.
We observed that the accordance between $F_M^{(2)}(a)$ and $^m\! F_M^{(2)}(a)$ increased with increasing $\Lambda$,
and the marked region in Fig.~\ref{Fig13} is reduced.

\section{Conclusions}

In this paper we demonstrated that the pion-deuteron scattering amplitude $F_{\pi d}$,
calculated in the Fixed Centers Approximation according to Eq.~(\ref{1_2}),
is correct and can be used in a rather wide region of the complex scattering length $a$.
Simultaneously, we proved that the coordinate representation in the FCA (\ref{1_2})
is getting incorrect and so cannot be used in a narrow region surrounding the real positive $a$ half-axis.
This region is marked in Fig.~\ref{Fig13}. For $a$ belonging to this region the interaction
between the incident light particle and the pair of the fixed centers, while being attractive,
becomes anomalously strong. Due to this anomalously strong attraction, an infinite set of $\pi NN$ bound states is formed.

An analogous situation, where an infinite set of bound states emerges as well, occurs in
the problem of the motion of a particle in a singular attractive potential, e.g.,

$$V(r)=-\frac{\beta}{r^n},$$
with $\beta >0$ and $n>2$, see, e.g., \cite{LL}, \S\S18, 35. The discrete part of the spectrum for
this problem is unbounded from below, i.e., the binding energy of the ground state is
infinite, whereas the size $r_0$ of the ground state goes to
zero. This is the phenomenon of ``the falling down to the center''.

Vitaly Efimov \cite{Efimov} showed in 1970 that the three-body system of identical bosons in the
limit of large two-body scattering length $a$ displays a specific discrete scale invariance. As
a consequence, there is an infinite number of 3-body bound states
and the ratio of binding energies of two successive 3-body bound states is approximately 515.
This cyclic behavior of the 3-body spectrum in the limit $a \rightarrow\infty$ has been rederived as the
leading order prediction of an EFT, see, e.g., \cite{Bedaque}. Note that the Efimov effect was
demonstrated to survive also in a model consisting of two heavy particles and a light one
when the light-heavy interaction leads to a zero-energy two-body bound state, see \cite{Fonseca}.

In our case, we found a similar (but not identical) cyclic behavior of the
spectrum of the system of a light particle interacting with a pair of fixed
centers, arising when the scattering length $a$, corresponding to the scattering
of the light particle on each of the fixed centers, is positive.

To study this problem,
we developed a method of getting the FCA amplitude by summing the 
MSS in the momentum space.
In this representation we solved the problem numerically. To compare the results of both methods, we introduced
a finite range of interaction $r_0=M^{-1}$ for each of the scatterers.
The point-like limit corresponds to the case $M\rightarrow\infty$. The solutions of Eqs.~(\ref{2_5}) and (\ref{3_3})
for positive values of the scattering length $a$ confirmed the existence of an infinite set of bound states in
the $(\pi NN)$-system. Along with that, we found that no unique solution for the
$\pi d$-scattering length exists in the momentum representation for the case of $a>0$. The function $^m\! F_M^{(2)}(a)$,
which represents the $\pi d$ scattering length as a function of $a$ and $M$, is an oscillating function of $M$
and hence gives no well-defined answer in the point-like limit $M\rightarrow\infty$.
We conclude therefore that the $\pi d$ scattering lenght is not well-defined for positive values of $a$.

It is also instructive to discuss expression (\ref{1_3}) for the scattering amplitude $F^{(2)}(a)$.
As seen from Eq.~(\ref{1_3}), $F^{(2)}(a)$ is complex even for real values of $a$.
At the same time the imaginary part of $F_M^{(2)}(a)$, calculated in the coordinate space according to Eq.~(\ref{1_3})
even with extremely small values of the imaginary part of $a$, is also non negligible, see, e.g., Fig.~\ref{Fig09}.
Obviously, expressions (\ref{1_2}) and (\ref{1_3}) violate unitarity!
This violation of unitarity is a direct consequence of the singularity of the interaction we
are dealing with. In the case of a singular attractive potential,
due to the phenomenon of ``the falling down to the center'',
there is a sink of particles and hence the unitarity violation.
We have a similar picture in the FCA: the interaction is getting supersingular, and the unitarity is violated.

Still, in the physically interesting cases like the $K^-d$ scattering
the imaginary part of the elementary ($K^-N$ in the example at hand) scattering length is comparable with the real
part. In this case the interaction in the 3-body system is not singular any more. As shown in Section 4,
formula (\ref{1_2}) is applicable in this case and both the coordinate and momentum space approaches to the 3-body scattering problem
give practically the same answers for the scattering amplitude $F^{(2)}(a)$.

It is interesting, however, to understand better the physical interpretation of this
``unphysical'' imaginary part. In hadronic atoms the imaginary part of the scattering length is
usually proportional to the width $\Gamma_{nl}$ of the atomic level. In the absence of a phyiscal absorption
it is difficult to say what this additional width means.

In the end it is worth to stress that the problem we studied here is a special case
of the general 3-body scattering problem, which is usually formulated in terms of the Faddeev equations.
We took the elementary $\pi N$-amplitudes to be constant and found that for positive and real scattering lengths $a$
the 3-body system has an infinite number of bound states. The introduction of a formfactor (or a finite range of the
interaction) does not resolve the problem completely: for values of $M$ larger than some critical value $M_0$, $M>M_0$, the system also
has bound states.

To solve this problem, we first need to understand what are the minor modifications of the elementary amplitudes that
will allow us to avoid the singularity that arises in the FCA with constant amplitudes or to decipher its nature.

\section*{Acknowledgments}

\begin{sloppypar}
We thank V.~E.~Tarasov and V.~G.~Ksenzov for useful discussions, C.~Hanhart and E.~Epelbaum
for their interest to our work.
We are extremely grateful to V.~A.~Lensky for careful reading of our manuscript and for
many useful questions and remarks. We are also thankful to V.~V.~Baru for useful and constructive criticism.
This work was supported partly by the grant NSh-3172.2012.2. One of the authors (AEK) thanks also the DFG and NSFC
grant CNC 110 for partial support.
\end{sloppypar}

\end{document}